# High resolution, High contrast optical interface for defect qubits

Jong Sung Moon,[1,†] Haneul Lee,[1,†] Jin Hee Lee,[1] Woong Bae Jeon,[1] Dowon Lee,[1] Junghyun Lee,[2] Seoyoung Paik,[2] Sang-Wook Han,[2] Rolf Reuter,[3] Andrej Denisenko,[3] Joerg Wrachtrup,[3] Sang-Yun Lee,[2,4] and Je-Hyung Kim[1]*

[1]School of Nature Science, Department of Physics, Ulsan National Institute of Science and Technology (UNIST), Ulsan 44919, Republic of Korea

[2]Center for Quantum Information, Korea Institute of Science and Technology, Seoul 02792, Republic of Korea

[3]Physikalisches Institut and Research Center SCOPE and Integrated Quantum Science and Technology (IQST), University of Stuttgart, Pfaffenwaldring 57, 70569 Stuttgart, Germany

[4]Department of Physics and Photon Science, Gwangju Institute of Science and Technology, Gwangju 61005, Republic of Korea

**Point defects in crystals provide important building blocks for quantum applications. To initialize, control, and read-out their quantum states, an efficient optical interface for addressing defects with photons is required. However, conventional confocal fluorescence microscopy with high refractive index crystals has limited photon collection efficiency and spatial resolution. Here, we demonstrate high resolution, high contrast imaging for defects qubits using microsphere-assisted confocal microscopy. A microsphere provides an excellent optical interface for point defects with a magnified virtual image that improves spatial resolution up to ~$\lambda/5$ as well as an optical signal-to-noise ratio by four times. These features enable individual optical addressing of single photons and single spins of spatially-unresolved defects in conventional confocal microscopy with improved signal contrast. The combined optical tweezers show the possibility of positioning or scanning the microspheres for deterministic coupling and wide-field imaging of defects. The approach does not require any complicated fabrication and additional optical system but uses simple micro-optics off-the-shelf. From these distinctive advantages of the microspheres, our approach can provide an efficient way for imaging and addressing closely-spaced defects with higher resolution and sensitivity.**

**Introduction**

Optically active point defects in crystals such as nitrogen- or silicon-vacancy centers in diamond have emerged as the most attractive candidates for implementing quantum technologies in solid-state platforms. These defects can serve single photons and a long coherent spin over 1 ms at room temperatures[1, 2]. Beyond the use of single defects, recent efforts for scaling up the quantum system have demonstrated multi-qubit quantum resistors by connecting multiple electrons as well as nuclear spins[3, 4], superradiant emission from multiple resonant emitters[5, 6], and wide-field imaging of magnetic or electric fields[7-9]. Therefore, optically addressing multiple adjacent defects is of paramount importance and requires an optical interface that resolves individual defects with high resolution and high readout contrast. However, accessing the defects in bulk crystals using conventional optical microscopy lacks such capabilities due to diffraction-limited spatial resolution and background fluorescence from crystals, substrates, and other defects.

There have been lots of efforts for improving the optical interface by fabricating micro/nanophotonic structures such as nanowires[10], nanobeams[11], bulls eyes[12], and solid-immersion lenses[13-15]. These structures have dramatically improved the light extraction efficiency more than an order of magnitude and enhanced the radiative recombination rate by the Purcell effect[16]. However, these approaches require a sophisticated fabrication process for materials with high hardness such as diamond, and the nanoscale devices often aggravate decoherence of optical and spin qubits[17-19]. Furthermore, to precisely couple the defects with photonic structures at the maximum optical mode, the lithography with patterned masks or site-selective generation of defects are required[20, 21].

On the other hand, the spatial resolution can be improved using various super-resolution techniques such as near-field probes[22], plasmon gratings[23], stimulated emission depletion microscopy[24, 25], and Fourier magnetic imaging[26]. Although these techniques have led to a significant improvement in the spatial resolution less than 10 nm, they are still not widely used in the defect's imaging due to several limitations: limited imaging depth, low optical throughput, non-radiative loss, and complicated imaging systems combined with a depletion beam or a gradient magnetic field. Therefore, addressing closely-spaced defects and their interaction with high spatial resolution and high readout contrast still remains a challenge.

Here, we introduce a high resolution, high contrast optical interface for defects qubits based on hybrid integration of dielectric micro-optics. Dielectric microspheres have several advantages over previous approaches using fabrication-based photonic structures. First of all, the microspheres are off-the-shelf optical elements with wide selection ranges in their materials and sizes. Simple dispersion of the microsphere on the crystals couples them to the defects and enhances the light extraction. The second advantage is the possible super-resolution imaging beyond the diffraction limit by sub-diffraction-limited illumination and far-field propagation of evanescent waves[27-30]. In particular, the microspheres form virtual or real images far apart from the actual sample surface[31]. This spatially translated image separates the defects qubits from optical background noise existing near the sample surface. The capability of high-resolution imaging with low background signals enables us to resolve multiple adjacent defects and optically address the photonic and spin qubits of those defects independently. Furthermore, we also combine the optical tweezers that trap and move the microspheres for future deterministic positioning and wide-field scanning defects. Therefore, together with fabrication-less simplicity, the microspheres provide a relocatable efficient optical interface for defect qubits.

## Result

We prepared point defects in diamond by an ion-implantation of $N_2$ ions that creates randomly distributed nitrogen-vacancy (NV) color centers near the surface within 100 nm (See Materials and Methods).

For microsphere-assisted confocal fluorescence imaging of defects in diamond, we adopt $BaTiO_3$ microspheres, having a high refractive index ($n$ ~1.9-2.1) and a diameter of 10~20 μm. The use of the high-refractive-index spheres is more beneficial for the defects in diamond ($n$~2.42) than other low index microspheres such as soda-lime glass ($n$~1.51), polystyrene ($n$~1.59) since the small refractive index contrast between the microsphere and diamond can minimize the total internal reflection at the interface. The microsphere also has another interface with the surrounding environment, and the relative refractive index contrast with the environment determines the light propagation through the microsphere. A small (large) index contrast with the environment forms a virtual (real) image below (above) the sample[31]. To maintain the advantage of high-index microspheres but reduce the index contrast with the environment, we immersed the microspheres in immersion oil (*n*=1.518). By doing so, we can bring the capability of super-resolution imaging to a virtual image plane[29-31]. Furthermore, the small index contrast between the spheres and immersion oil enables us to trap the microsphere using optical tweezers, which is shown later.

For fluorescence imaging, we used a home-built confocal microscopy setup at room temperature. We excited the sample with 532 nm continuous-wave laser and collected the fluorescence from the sample with an objective lens in immersion oil environment (NA=1.4). Then, we spatially filtered the sample emission with a 50 μm pin-hole and sent it to avalanche photodiodes. (see Materials and Methods)

Figure 1(a) illustrates the microsphere-assisted microscopy for imaging defects in a diamond. Immersing the high-refractive-index microsphere in a liquid (*n*~1.518) forms a virtual image of defects underneath the sample surface. We first vertically scans the sample to determine image planes with and without a microsphere lens. As shown in Fig. 1(b), outside the microsphere, the confocal map displays the fluorescence from defects at $z = 0$ near the diamond surface. In comparison, the defects under the microsphere form the virtual image at a lower plane near $z = -12$ μm. Figure 1(c, d) exhibits x-y lateral maps at two different image planes of $z = 0$ and $z = -12$ μm. In comparison, the virtual image in Fig. 1(d) noticeably shows the defects with better contrast and resolution than those at the sample surface outside the microsphere. Such improvement in the image quality is relevant to the optical phenomena in the microsphere. As described in Fig. 1(e), a microsphere, as a spherical lens, creates a magnified virtual image. The magnification factor varies with the size of spheres, optical index contrast, and the distance of defects from the microsphere[27, 30, 32].

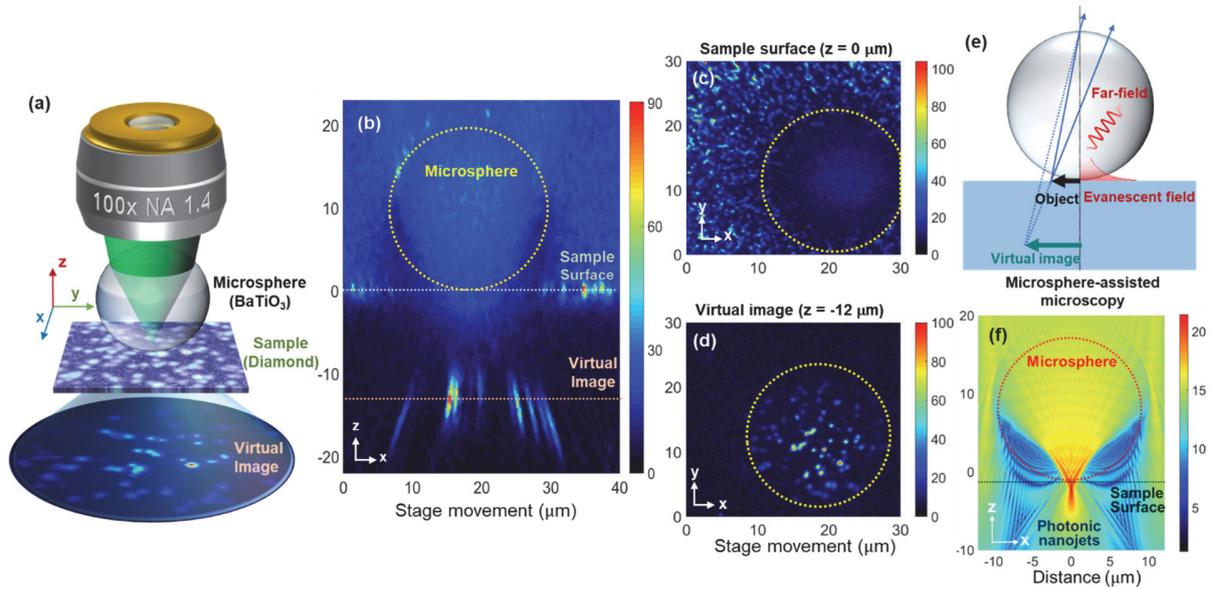

**Fig. 1** (a) Schematic description of microsphere-assisted microscopy that creates a virtual image underneath the sample surface. (b) Vertically scanned confocal image of defects with a microsphere. The dotted lines indicate the microsphere, sample surface, and virtual image plane. (c,d) x-y scanned confocal images with the microsphere at different vertical planes of z = 0 μm (sample surface) (c) and z = −12 μm (virtual image plane). (d) The color bar denotes the photoluminescence (PL) intensity in the unit of kcounts per second. (e) Schematic representation of optical phenomena in the microsphere that magnifies the image and couples the evanescent field. (f) Numerical simulation of an incident plane wave toward a 20 μm-size microsphere on a diamond. A logarithmic color map of the intensity ($|E|^2$) shows the photonic nanojet effect that tightly focuses the incident wave at the bottom of the microsphere.

More importantly, the microspheres improve the spatial resolution beyond the diffraction limit in conventional confocal microscopy[27-30, 33]. Two optical phenomena in microspheres play an important role in super-resolution imaging. One is near-field coupling within the gap (λ/2) from the surface as described in Fig. 1(e), and the other one is sub-diffraction-limited illumination, so-called photonic-nanojet[30, 33-35]. Figure 1(f) shows a finite-difference time-domain simulation of cross-sectional $|E|^2$ profiles with a 20 μm-diameter microsphere. The dielectric microsphere induces constructive interference between diffracted waves[30] and tightly focuses the incident wave into a localized spot with a beam waist of 232 nm for the illuminating wavelength of 532 nm. In addition, the microsphere couple high-frequency evanescent waves to far-field emission near the contact area around a quarter of the sphere size[36], and therefore, it improves the spatial resolution further. These effects on the excitation and collection enhance the spatial resolution of the microscopy up to $\lambda/7$ and have demonstrated the super-resolution capability for 50-nm-size nanoparticles[29] and 75-nm-size virus samples[27].

In Figure 2(a, b), we directly compare the images for the same defects with and without the microsphere. We scan the lateral area with the same scan range for the comparison, but the microsphere shows the magnified virtual image that is almost doubled. We determine the magnification factor ($M$) of the virtual images from the non-magnified confocal image without the microsphere. Figure 2(c) shows the experimentally measured $M$ with depth. The magnification increases from 1.9 to 2.5 in the range of $z = -9 \sim -15$ μm. The magnification in the spherical lens system can be approximated using geometrical optics[27]. By considering a spherical aberration, the microsphere has a focal length $f(x) = \frac{x}{\sin[2\sin^{-1}(x/R) - 2\sin^{-1}(n_r^{-1}(x/R))]}$, where $x$ is the transverse distance from the optical axis, $R$ is the radius of a microsphere, and $n_r = \frac{n_{microsphere}}{n_{oil}} = 1.38$ is the refractive index contrast between the microsphere and the oil. Then the sphere creates a virtual image at a distance ($d_v = \frac{(R+\delta)f}{f - R - \delta} - R$) from the sample surface. The magnification is given as $M = \frac{R + d_v}{R + \delta}$, where $\delta$ is the depth of defects from the sample surface. Following the above formula with $R = 10$ μm, $\delta = 100$ nm, and $x = 1$ μm, we can calculate the virtual image plane at $-13$ μm with $M = 2.28$. The magnification will vary with the observed virtual plane as $M(z) = \frac{R - z}{R + \delta}$. The theoretically calculated values are well matched to the experimentally observed results in Fig. 2(c).

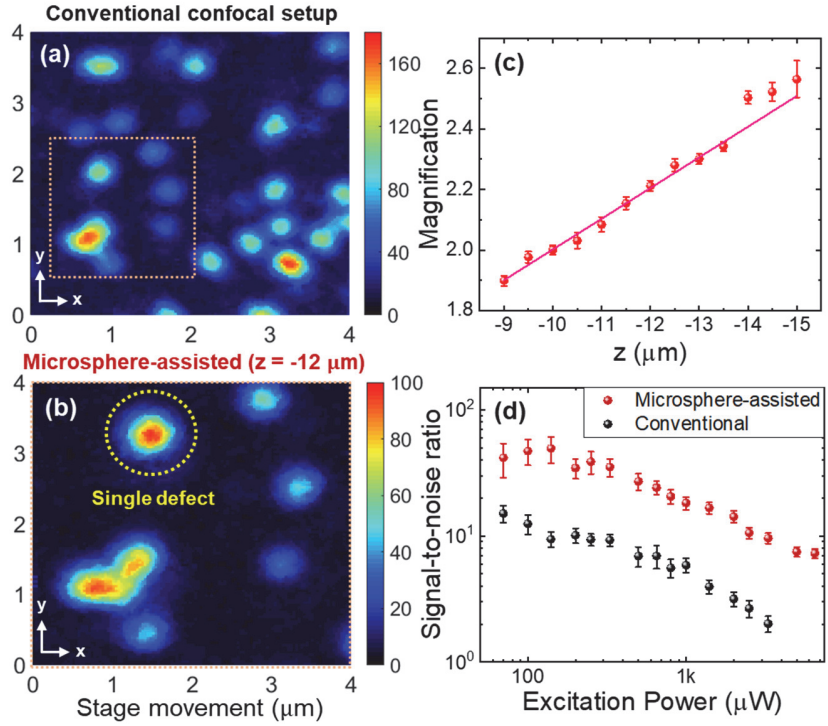

**Fig. 2** (a,b) Comparison of confocal fluorescence images of the conventional (a) and microsphere-assisted (b) microscopy. Two images are laterally scanned in the same x-y scan range at $z = 0$ (without microsphere) and $z = -12$ μm (with microsphere). The magnified image in (b) corresponds to the area, marked as the orange-dashed box in (a). (c) Magnification factors measured at different virtual image planes in the microsphere-assisted microscopy. A solid line denotes a calculated magnification factor with the observed virtual plane ($z$). (d) Comparison of the signal-to-noise ratio with (red) and without (black) the microsphere. The yellow-circle in (b) denotes the single defect used for the comparison in (d).

In terms of brightness, the high-index spheres can couple the evanescent waves near the diamond's surface to free space. From the numerical simulation, we calculate the improved extraction efficiency of dipoles under the microspheres by a factor of two. In the experiment, we observe the enhancement in brightness up to 40% by comparing the saturation intensity of the same defect with and without a microsphere. The discrepancy between the simulation and observation can be due to the narrow spatial window of the microsphere for the maximum brightness. The simulation shows that the maximum enhancement only occurs within about 1 μm right under the sphere. This window can be wider by increasing the size of the microsphere.

Although the enhancement in the brightness is moderate, the microsphere enables a large increase in a signal-to-noise ratio with substantially reduced background fluorescence in the virtual image compared to that at the sample surface. Increasing the signal-to-noise ratio is, in particular, important for quantum applications since they manipulate the information at the single-photon and single-spin levels. However, the optical excitation commonly produces unwanted background fluorescence from the scattered photons and autofluorescence from the environment, such as immersion oil, hosting crystal, and nearby defects. As we can observe in Fig. 1(b), the background fluorescence is relatively brighter above the sample surface than inside the diamond, where the virtual image is measured. Therefore, the spatial separation of the virtual image far apart from the sample surface is beneficial to avoid the optical background noise. Figure 2(d) compares the signal-to-noise ratio of the same single defect (yellow circle in Fig. 2(b)) with and without the microsphere. The result shows a large improvement in the signal-to-noise ratio by an average of four times for the virtual image.

The fluorescence of a single atomic defect can serve as a perfect point-spread function for measuring the spatial resolution of the image. To determine the enhancement in spatial resolution in the microsphere-assisted microscopy, we calibrate the magnified virtual images with the measured $M$ in Fig. 2(c). After the calibration, we compare the spatial resolution of the images for the same defects with and without the microsphere in Figure 3(a, b). The spatially-unresolved defect groups D and T in the conventional microscopy are well resolved as separated defects under the microsphere, labels as D1-D2 and T1-T3. The result clearly shows the improved resolving power in the microsphere-assisted microscopy. For measuring the full width at half maximum

(FWHM) of the point-spread function, we select the single defect labeled as S in Fig. 3(a, b) and plot its cross-sectional intensity curve in Fig. 3(c). In conventional microscopy, the FWHM of the fitted Gaussian function is about 280 nm. Considering the center wavelength of 700 nm for the broad spectrum (600 ~ 800 nm) of NV centers, the value corresponds to a spatial resolution of ~$\lambda/2.5$, close to the Abbe diffraction limit of $\lambda/(2*\mathrm{NA})$, where the NA is 1.4 with an oil immersion lens. For the microsphere-assisted microscopy, we measure the point-spread function at two different virtual image planes at $z = -13$ and $z = -17$ μm and achieve the FWHM of 188 nm ($\lambda/3.7$) and 142 nm ($\lambda/5$), respectively. Therefore, the microsphere-assisted microscopy serves the spatial resolution beyond the diffraction limit in conventional microscopy. Fig. 3(d) plots the spatial resolution as well as the intensity with the imaging depth from the surface. The microsphere forms the best focal image at around $z = -12 \sim -13$ μm with the maximum intensity. As the image plane goes below the best focal plane, we observe higher resolution but lose the intensity. At $z = -17$ μm, the intensity is reduced to 20% compared to the maximum intensity. Also, the intensity profile at $z = -17$ μm in Fig. 3(c) shows two shoulder peaks. These peaks originate from airy patterns, which become stronger as we pass the focal plane.

From the enhanced resolving power and improved optical contrast, we examine the closely-spaced defects, not resolved in the conventional microscopy. In Fig. 3(e, f), the conventional microscopy shows the intensity profile of the defect group D as a merged single peak, whereas the microsphere-assisted microscopy shows two separated peaks for the defects D1 and D2 in Fig. 3(b). Each single quantum emitter behaves as a single-photon source, quantified with an antibunching dip, ideally $g^{(2)}(0) = 0$. However, as multiple defects are spatially and spectrally overlapped, the value increases as $g^{(2)}(0) = 1 - \frac{1}{N}$, where $N$ is the number of quantum emitters. For investigating single-photon purity, we measure the second-order photon correlation $g^{(2)}(\tau)$ using Hanbury Brown and Twiss measurements and fit the data using the following equation, $g^{(2)}(\tau) = 1 - \left(1 - g^{(2)}(0)\right)e^{-\frac{|\tau|}{\tau_c}}$, where $\tau_c$ is a time constant relevant to the transitions between the ground and excited states. The single defect (S) shows expected low $g^{(2)}(0)$ values of $0.08 \pm 0.05$ ($0.06 \pm 0.04$) in the conventional (microsphere-assisted) microscopy. Then we measure the $g^{(2)}(\tau)$ for the closely-spaced two defects. For the defect group D, we record the histogram at the maximum intensity of the merged peak and achieve $g^{(2)}(0) = 0.23 \pm 0.07$ as shown in Fig. 3(h). The value is noticeably increased than that from the single defect. On the other hand, for spatially-resolved single defects D1 and D2, the $g^{(2)}(0)$ values of each single defect D1 and D2 in the virtual image remain as low as $0.03 \pm 0.04$ and $0.09 \pm 0.04$, respectively (Fig. 3(h)). Such high single-photon purity validates that the magnified, high-resolution imaging enables us to optically isolate single photons from multiple adjacent defects independently. It is also worth noting that $g^{(2)}(0)$ value is still below 0.5 for spatially-overlapped two defects in the conventional microscopy. The emission from two independent, identical quantum emitters are expected to have $g^{(2)}(0) \sim 0.5$. However, even with two defects, they can have the $g^{(2)}(0)$ value below 0.5 when the brightness of each emitter is not identical or the measurement is spatially one-sided between two defects. Besides, these defects are temporally blinking, making them behave as a mixture of single and two emitters at the given experimental conditions. The fact $g^{(2)}(0) < 0.5$ is often used as a criterion of a single defect, but the result suggests that $g^{(2)}(0) < 0.5$ alone cannot strongly support the single quantum emitters. Therefore, having a high-resolution imaging system will be much more beneficial to identify the single defects and independently manipulate closely-spaced defects.

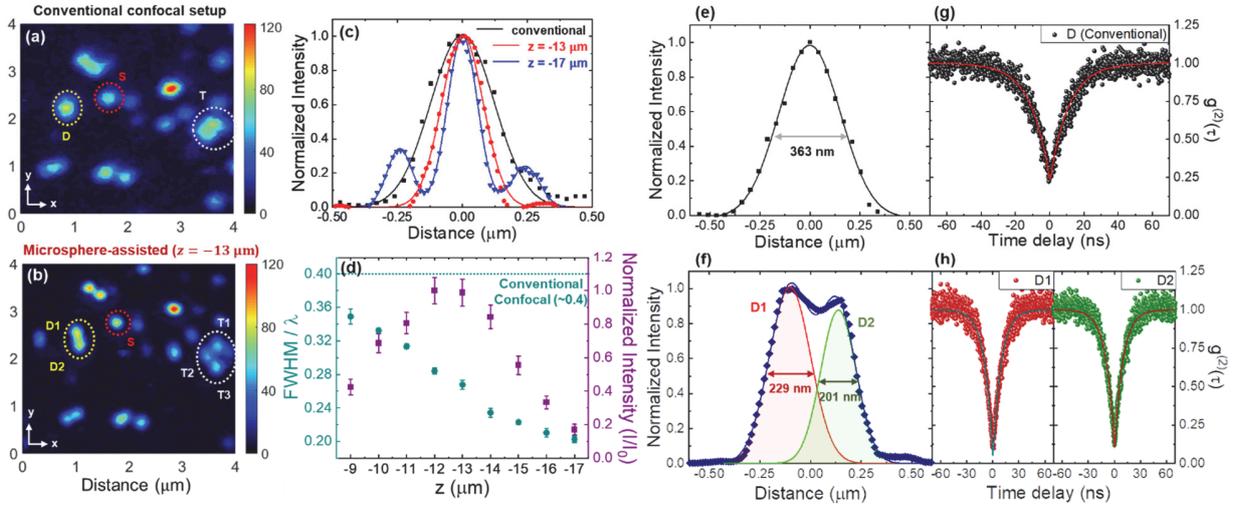

**Fig. 3** (a, b) Confocal images of defects in conventional (a) and microsphere-assisted (b) microscopy. The virtual image at $z = -13$ μm in (b) is rescaled with the magnification factor for comparing with the conventional confocal image in (a). Defects labeled as S, D, and T denote single, double, and triple defects, which are merged in (a). These defects are well isolated and labeled (D1-2) and (T1-3) in (b). (c) Cross-sectional intensity profiles of defect S from the conventional confocal microscopy (black dots) and microsphere-assisted microscopy at different z positions of $z = -13$ (red dots) and $-17$ μm (blue dots). Solid lines are fitted lines with Gaussian functions. (d) Spatial resolution and normalized intensity of the single defect S as a function of the z position of the virtual image. A green-dashed line indicates the diffraction limit of conventional microscopy. (e,f) Intensity profiles of two adjacent defects D in (a) and D1-2 in (b). The defects are not resolved in (e), while they are resolved in (f). The data in (e), (f) are fitted with single and two-separated Gaussian functions, respectively. (g,h) Second-order photon correlation curves, measured at the maximum peak in D (g) and each peak at D1-2 (h). The red lines are fitted $g^{(2)}(\tau)$ curves.

Next, we demonstrate that the microsphere-assisted microscopy improves the spin readout contrast and enables us to address single spins of the adjacent defects independently. The NV centers in diamond have a zero-field splitting of 2.87 GHz[37] without an external magnetic field, which can be detected by an optically detected magnetic resonance (ODMR) measurement. Figure 4(a,b) show the ODMR spectra of the same single defect in conventional and microsphere-assisted microscopy. In comparison, the single defect under the microsphere increases the ODMR contrast from $15 \pm 1.1\%$ to $25 \pm 0.9\%$. We attribute this increase to the improved signal-to-noise ratio in the virtual image. To prove that we can address single spins of the unresolved defects in a conventional optical diffraction limit, we examine the ODMR spectrum of two closely-spaced defects using both microscopy setups. These two defects are spatially overlapped in the conventional microscopy, as shown in the inset of Fig. 4(c). To confirm that they are two defects, we apply a magnetic field (~50 G) that breaks the degeneracy of S=1. Since two defects have different spin orientations to the applied magnetic field, they have a different amount of Zeeman splitting. In Fig. 4(c), we observe four peaks from two defects in the ODMR spectrum. Each defect shows a splitting $\delta_1 = \pm 50$ MHz and $\delta_2 = \pm 118$ MHz, respectively, from the center at 2.87 GHz. Since these defects are within the diffraction limit of the conventional microscopy, we cannot optically readout each spin's state. However, with the microsphere-assisted confocal microscopy, these defects are clearly resolved, as shown in the inset of Fig. 4(d). We label these defects as 1 and 2, then readout the ODMR signals from each defect as shown in Fig. 4(d) and (e), respectively. In contrast to the previous four peaks from the unresolved two defects, each ODMR signal shows only two separated peaks from each defect ($\delta_1 = \pm 55$ MHz for defect 1 and $\delta_2 = \pm 120$ MHz for defect 2) without crosstalk from another defect. The difference in $\delta_{1,2}$ between the setups is due to the slight change in the alignment of the magnet during the measurements. Therefore, the microsphere-assisted imaging with enhanced spatial resolution and signal contrast can lead to individual addressing of both single photons and spins where defects are spatially-unresolved in the conventional microscopy.

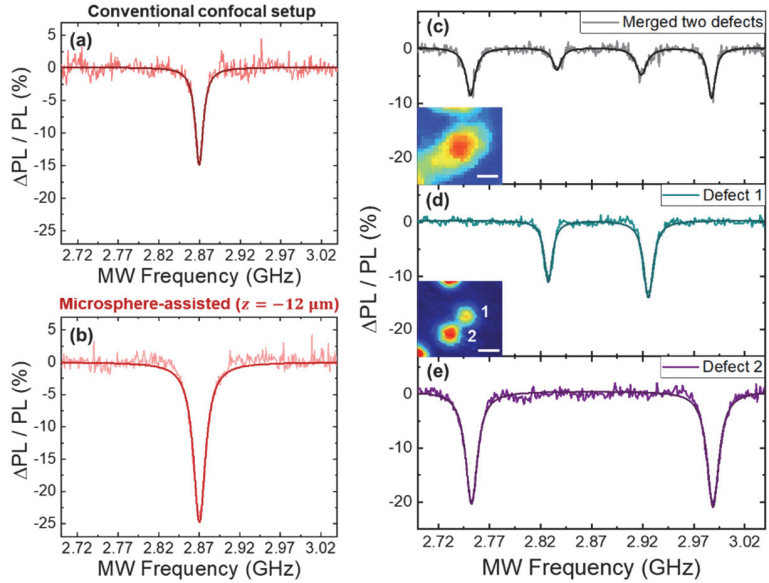

**Fig. 4** (a, b) Optically detected magnetic resonance (ODMR) spectra of the single defect with and without a microsphere. (c-e) ODMR spectra with a magnetic field of about 50 G for spatially-unresolved two defects (c) in the conventional confocal microscopy and spatially-resolved defect 1 (d) and 2 (e) in the microsphere-assisted microscopy. The inset images represent confocal mapping images of the same defects in the conventional and microsphere-assisted microscopy. Scale bars in the confocal maps denote 200 nm.

Finally, we show the distinct advantage of a microsphere lens as a relocatable optical interface. The position of the microspheres can be easily controlled by mechanical[38, 39] or optical forces[40, 41]. Here, we combine optical tweezers for position control of the microsphere. Generally, high-index microspheres are difficult to trap using optical tweezes due to a large scattering force at the interface[42]. However, the immersion oil lowers the optical index contrast between the microsphere and the environment, enabling us to overcome the van der Waals and scattering forces. The images in Fig. 5(a, b) exhibits the trapping and relocating of individual microspheres with the optical tweezers. We note that the microspheres are attached to the substrate without the trapping laser. Therefore, once we position the spheres, they stay in place even after a few days.

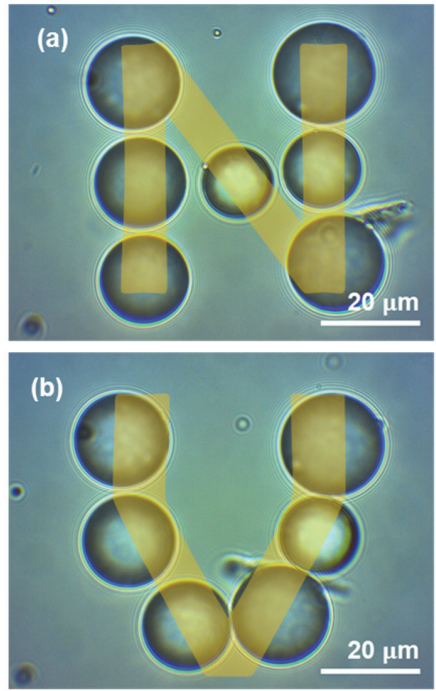

**Fig.5** (a,b) Trapping and positioning of microspheres in immersion oil with optical tweezers. The microspheres are rearranged to represent alphabets N (a) and V (b).

## Discussion

In conclusion, we have demonstrated the high resolution and high contrast optical interface that independently addresses single photons and single spins of merged defects in the conventional microscopy. The microsphere showed great potential as a simple and effective optical interface between the point defects and photons with a sub-diffraction-limited spatial resolution up to ~$\lambda/5$ and an improved signal-to-noise ratio of more than four times compared to that of the conventional confocal microscopy. The nanoscale spatial resolution with enhanced sensitivity is a key requirement for nanoscale imaging of magnetic fields and temperature[7, 43, 44]. Also, the capability to address adjacent defects independently is essential to manipulate multiple defects and their interactions. Depending on the types of interaction, the required interacting distance will vary.

For example, photon-mediated collective interaction such as superradiance occurs in the wavelength scale[5, 6, 45], sensing the electric or magnetic fields requires the distance in the range of 100 nm[9], and the dipole-dipole interaction occurs at a closer distance of sub-100 nm. The possible limit of the resolution of the microsphere-assisted microscopy can be improved further by changing the size and index contrast of the microspheres. For the small-size spheres less than 10 μm, the photonic nanojet will be formed near the surface with the smallest waist and govern the resolution[33], while the large-size spheres are advantageous for coupling more evanescent fields in a wide field of view[27]. A 100 μm-sized microsphere has shown a resolution of $\sim\lambda/7$ that resolves 75 nm-sized features[27], and combining plasmonic structures can strengthen the evanescent field, increasing the resolution up to $\sim\lambda/10$[41]. Resonantly driving whispering gallery modes existing in the spheres also further enhances the resolutions[32]. Therefore, the capability of super-resolution imaging with microspheres would reach the region where various interactions exist between multiple defects.

The integration of microspheres does not use any fabrication process, so it would not create additional sources of dephasing, which takes place with the nanostructure-based optical interfaces. Instead, our approach can be simply adapted with various defects in bulk crystals and conventional microscopy systems. Furthermore, the possibility of relocating these micro-optics extends their uses for scanning and deterministic positioning them for the defects. Therefore, the microsphere-assisted microscopy will pave a new way of exploring solid-state quantum systems.

**Materials and Methods.**

*Sample preparation*

We prepared nitrogen-vacancy (NV) color centers in diamond (Element Six) by an ion-implantation (40 keV $N_2$ ions) followed by thermal annealing at 950 °C. The sample was cleaned in a boiling mixture of Sulfuric acid, Nitric acid, and Perchloric acid to remove a graphite layer on the surface. The process formed randomly distributed NV color centers near the surface within 100 nm with a density of 1~2 $\mu m^{-2}$. To integrate the $BaTiO_3$ microspheres (Cospheric BTGMS-4.15) on the diamond sample, we diluted them in isopropyl alcohol and dispersed on the sample surface. For the sake of reducing the index contrast and bringing the capability of an optical trap, we immersed the high-index spheres in immersion oil (Carl Zeiss Immersol$^{TM}$ 518F, $n$=1.518).

*Optical setup*

For fluorescence imaging, we used a home-built confocal microscopy setup. A 532 nm continuous-wave excitation laser excited the defects in the diamond on a three-axis piezoelectric stage. An objective lens (Olympus UPlanSApo, 100x, NA=1.4) focused the excitation laser and collected the fluorescence signal from defects. We spectrally filtered out the exciting laser by using a dichroic mirror and a long pass filter and then spatially filtered out the emission by using a 50 μm-pinhole and two lenses, followed by a fiber-coupled silicon single-photon avalanche photodiode (Excelitas, SPCM-AQRH). To identify single-photon emission, we performed a Hanbury Brown and Twiss experiment with a 50:50 fiber beamsplitter coupled with two silicon single-photon avalanche photodiodes. To optically trap the microsphere, we used a 60 mW, 820 nm continuous-wave laser and focused the laser using a high numerical aperture (NA) objective lens (Carl Zeiss EC Epiplan-Neofluar 100x, NA=0.9). For optically detected magnetic resonance, we applied a microwave using a thin gold wire (Ø~40 μm), which is connected to a RF generator (R&S, SMBV100A) followed by an amplifier (Mini-Circuits, ZHL-15W-422-S+).

**Corresponding Author**

*E-mail: jehyungkim@unist.ac.kr

**Author Contributions**

†These authors contributed equally to this paper.

ACKNOWLEDGEMENT

This work is supported by the National Research Foundation of Korea (MSIT) (NRF-2019M3E4A1078664, NRF-2020M3H3A1098869); Institute of Information and Communications Technology Planning and Evaluation (IITP) grant (2019-0-00434); the ITRC (Information Technology Research Center) support program (IITP-2020-0-01606) supervised by the IITP; the KIST Institutional Program (2E29580-19-146);

REFERENCE

[1] G. Balasubramanian, P. Neumann, D. Twitchen, M. Markham, R. Kolesov, N. Mizuochi, J. Isoya, J. Achard, J. Beck, J. Tissler, V. Jacques, P. R. Hemmer, F. Jelezko, and J. Wrachtrup, "Ultralong spin coherence time in isotopically engineered diamond,"  Nat. Mater. **8**, 383-387 (2009)

[2] T. Ishikawa, K.-M. C. Fu, C. Santori, V. M. Acosta, R. G. Beausoleil, H. Watanabe, S. Shikata, and K. M. Itoh, "Optical and Spin Coherence Properties of Nitrogen-Vacancy Centers Placed in a 100 nm Thick Isotopically Purified Diamond Layer,"  Nano Lett. **12**, 2083-2087 (2012)

[3] P. Neumann, R. Kolesov, B. Naydenov, J. Beck, F. Rempp, M. Steiner, V. Jacques, G. Balasubramanian, M. L. Markham, D. J. Twitchen, S. Pezzagna, J. Meijer, J. Twamley, F. Jelezko, and J. Wrachtrup, "Quantum register based on coupled electron spins in a room-temperature solid,"  Nature Physics **6**, 249-253 (2010)

[4] C. E. Bradley, J. Randall, M. H. Abobeih, R. C. Berrevoets, M. J. Degen, M. A. Bakker, M. Markham, D. J. Twitchen, and T. H. Taminiau, "A Ten-Qubit Solid-State Spin Register with Quantum Memory up to One Minute,"  Physical Review X **9**, 031045 (2019)

[5] A. Angerer, K. Streltsov, T. Astner, S. Putz, H. Sumiya, S. Onoda, J. Isoya, W. J. Munro, K. Nemoto, J. Schmiedmayer, and J. Majer, "Superradiant emission from colour centres in diamond,"  Nature Physics **14**, 1168-1172 (2018)

[6] C. Bradac, M. T. Johnsson, M. v. Breugel, B. Q. Baragiola, R. Martin, M. L. Juan, G. K. Brennen, and T. Volz, "Room-temperature spontaneous superradiance from single diamond nanocrystals,"  Nature

Communications **8**, 1205 (2017)

[7] C. Foy, L. Zhang, M. E. Trusheim, K. R. Bagnall, M. Walsh, E. N. Wang, and D. R. Englund, "Wide-Field Magnetic Field and Temperature Imaging Using Nanoscale Quantum Sensors," ACS Appl. Mater. Interfaces **12**, 26525-26533 (2020)

[8] K. Mizuno, H. Ishiwata, Y. Masuyama, T. Iwasaki, and M. Hatano, "Simultaneous wide-field imaging of phase and magnitude of AC magnetic signal using diamond quantum magnetometry," Scientific Reports **10**, 11611 (2020)

[9] F. Dolde, H. Fedder, M. W. Doherty, T. Nöbauer, F. Rempp, G. Balasubramanian, T. Wolf, F. Reinhard, L. C. L. Hollenberg, F. Jelezko, and J. Wrachtrup, "Electric-field sensing using single diamond spins," Nature Physics **7**, 459-463 (2011)

[10] T. M. Babinec, B. J. M. Hausmann, M. Khan, Y. Zhang, J. R. Maze, P. R. Hemmer, and M. Lončar, "A diamond nanowire single-photon source," Nat. Nanotechnol. **5**, 195-199 (2010)

[11] B. J. M. Hausmann, B. J. Shields, Q. Quan, Y. Chu, N. P. de Leon, R. Evans, M. J. Burek, A. S. Zibrov, M. Markham, D. J. Twitchen, H. Park, M. D. Lukin, and M. Lončar, "Coupling of NV Centers to Photonic Crystal Nanobeams in Diamond," Nano Lett. **13**, 5791-5796 (2013)

[12] L. Li, E. H. Chen, J. Zheng, S. L. Mouradian, F. Dolde, T. Schröder, S. Karaveli, M. L. Markham, D. J. Twitchen, and D. Englund, "Efficient Photon Collection from a Nitrogen Vacancy Center in a Circular Bullseye Grating," Nano Lett. **15**, 1493-1497 (2015)

[13] M. Jamali, I. Gerhardt, M. Rezai, K. Frenner, H. Fedder, and J. Wrachtrup, "Microscopic diamond solid-immersion-lenses fabricated around single defect centers by focused ion beam milling," Rev. Sci. Instrum. **85**, 123703 (2014)

[14] L. Robledo, L. Childress, H. Bernien, B. Hensen, P. F. A. Alkemade, and R. Hanson, "High-fidelity projective read-out of a solid-state spin quantum register," Nature **477**, 574-578 (2011)

[15] P. Siyushev, F. Kaiser, V. Jacques, I. Gerhardt, S. Bischof, H. Fedder, J. Dodson, M. Markham, D. Twitchen, F. Jelezko, and J. Wrachtrup, "Monolithic diamond optics for single photon detection," Appl. Phys. Lett. **97**, 241902 (2010)

[16] W. L. Barnes, G. Björk, J. M. Gérard, P. Jonsson, J. A. E. Wasey, P. T. Worthing, and V. Zwiller, "Solid-state single photon sources: light collection strategies," Eur. Phys. J. D **18**, 197-210 (2002)

[17] C. Bradac, T. Gaebel, N. Naidoo, M. J. Sellars, J. Twamley, L. J. Brown, A. S. Barnard, T. Plakhotnik, A. V. Zvyagin, and J. R. Rabeau, "Observation and control of blinking nitrogen-vacancy centres in discrete nanodiamonds," Nature Nanotechnology **5**, 345-349 (2010)

[18] A. Faraon, C. Santori, Z. Huang, V. M. Acosta, and R. G. Beausoleil, "Coupling of Nitrogen-Vacancy Centers to Photonic Crystal Cavities in Monocrystalline Diamond," Physical Review Letters **109**, 033604 (2012)

[19] J. Tisler, G. Balasubramanian, B. Naydenov, R. Kolesov, B. Grotz, R. Reuter, J.-P. Boudou, P. A. Curmi, M. Sennour, A. Thorel, M. Börsch, K. Aulenbacher, R. Erdmann, P. R. Hemmer, F. Jelezko, and J. Wrachtrup, "Fluorescence and Spin Properties of Defects in Single Digit Nanodiamonds," ACS Nano **3**, 1959-1965 (2009)

[20] T. van der Sar, E. C. Heeres, G. M. Dmochowski, G. de Lange, L. Robledo, T. H. Oosterkamp, and R. Hanson, "Nanopositioning of a diamond nanocrystal containing a single nitrogen-vacancy defect center," Appl. Phys. Lett. **94**, 173104 (2009)

[21] R. Fukuda, P. Balasubramanian, I. Higashimata, G. Koike, T. Okada, R. Kagami, T. Teraji, S. Onoda, M. Haruyama, K. Yamada, M. Inaba, H. Yamano, F. M. Stürner, S. Schmitt, L. P. McGuinness, F. Jelezko, T. Ohshima, T. Shinada, H. Kawarada, W. Kada, O. Hanaizumi, T. Tanii, and J. Isoya, "Lithographically engineered shallow nitrogen-vacancy centers in diamond for external nuclear spin sensing," New J. Phys. **20**, 083029 (2018)

[22] E. Betzig, and R. J. Chichester, "Single Molecules Observed by Near-Field Scanning Optical Microscopy," Science **262**, 1422-1425 (1993)

[23] K. A. Willets, A. J. Wilson, V. Sundaresan, and P. B. Joshi, "Super-Resolution Imaging and Plasmonics," Chemical Reviews **117**, 7538-7582 (2017)

[24] S. Pezzagna, D. Wildanger, P. Mazarov, A. D. Wieck, Y. Sarov, I. Rangelow, B. Naydenov, F. Jelezko,

S. W. Hell, and J. Meijer, "Nanoscale Engineering and Optical Addressing of Single Spins in Diamond," Small **6**, 2117-2121 (2010)
[25] E. Rittweger, K. Y. Han, S. E. Irvine, C. Eggeling, and S. W. Hell, "STED microscopy reveals crystal colour centres with nanometric resolution," Nature Photonics **3**, 144-147 (2009)
[26] K. Arai, C. Belthangady, H. Zhang, N. Bar-Gill, S. J. DeVience, P. Cappellaro, A. Yacoby, and R. L. Walsworth, "Fourier magnetic imaging with nanoscale resolution and compressed sensing speed-up using electronic spins in diamond," Nature Nanotechnology **10**, 859-864 (2015)
[27] L. Li, W. Guo, Y. Yan, S. Lee, and T. Wang, "Label-free super-resolution imaging of adenoviruses by submerged microsphere optical nanoscopy," Light Sci. Appl. **2**, e104-e104 (2013)
[28] H. Zhu, W. Fan, S. Zhou, M. Chen, and L. Wu, "Polymer Colloidal Sphere-Based Hybrid Solid Immersion Lens for Optical Super-resolution Imaging," ACS Nano **10**, 9755-9761 (2016)
[29] Z. Wang, W. Guo, L. Li, B. Luk'yanchuk, A. Khan, Z. Liu, Z. Chen, and M. Hong, "Optical virtual imaging at 50 nm lateral resolution with a white-light nanoscope," Nature Communications **2**, 218 (2011)
[30] Z. Wang, and B. Luk'yanchuk, "Super-resolution imaging and microscopy by dielectric particle-lenses," in *Label-Free Super-Resolution Microscopy*(Springer, 2019), pp. 371-406.
[31] H. S. S. Lai, F. Wang, Y. Li, B. Jia, L. Liu, and W. J. Li, "Super-resolution real imaging in microsphere-assisted microscopy," PLoS One **11**, e0165194 (2016)
[32] A. V. Maslov, and V. N. Astratov, "Resolution and Reciprocity in Microspherical Nanoscopy: Point-Spread Function Versus Photonic Nanojets," Physical Review Applied **11**, 064004 (2019)
[33] H. Yang, R. Trouillon, G. Huszka, and M. A. M. Gijs, "Super-Resolution Imaging of a Dielectric Microsphere Is Governed by the Waist of Its Photonic Nanojet," Nano Lett. **16**, 4862-4870 (2016)
[34] Z. Chen, A. Taflove, and V. Backman, "Photonic nanojet enhancement of backscattering of light by nanoparticles: a potential novel visible-light ultramicroscopy technique," Optics Express **12**, 1214-1220 (2004)
[35] J. Zhu, and L. L. Goddard, "All-dielectric concentration of electromagnetic fields at the nanoscale: the role of photonic nanojets," Nanoscale Advances **1**, 4615-4643 (2019)
[36] A. Darafsheh, G. F. Walsh, L. Dal Negro, and V. N. Astratov, "Optical super-resolution by high-index liquid-immersed microspheres," Appl. Phys. Lett. **101**, 141128 (2012)
[37] A. Gruber, A. Dräbenstedt, C. Tietz, L. Fleury, J. Wrachtrup, and C. v. Borczyskowski, "Scanning Confocal Optical Microscopy and Magnetic Resonance on Single Defect Centers," Science **276**, 2012 (1997)
[38] F. Wang, L. Liu, H. Yu, Y. Wen, P. Yu, Z. Liu, Y. Wang, and W. J. Li, "Scanning superlens microscopy for non-invasive large field-of-view visible light nanoscale imaging," Nature Communications **7**, 13748 (2016)
[39] L. A. Krivitsky, J. J. Wang, Z. Wang, and B. Luk'yanchuk, "Locomotion of microspheres for super-resolution imaging," Sci. Rep. **3**, 3501 (2013)
[40] E. McLeod, and C. B. Arnold, "Subwavelength direct-write nanopatterning using optically trapped microspheres," Nat. Nanotechnol. **3**, 413-417 (2008)
[41] A. Bezryadina, J. Li, J. Zhao, A. Kothambawala, J. Ponsetto, E. Huang, J. Wang, and Z. Liu, "Localized plasmonic structured illumination microscopy with an optically trapped microlens," Nanoscale **9**, 14907-14912 (2017)
[42] X. Liu, S. Hu, Y. Tang, Z. Xie, J. Liu, and Y. He, "Selecting a Proper Microsphere to Combine Optical Trapping with Microsphere-Assisted Microscopy," Appl. Sci. **10** (2020)
[43] F. Casola, T. van der Sar, and A. Yacoby, "Probing condensed matter physics with magnetometry based on nitrogen-vacancy centres in diamond," Nature Reviews Materials **3**, 17088 (2018)
[44] S. Schmitt, T. Gefen, F. M. Stürner, T. Unden, G. Wolff, C. Müller, J. Scheuer, B. Naydenov, M. Markham, S. Pezzagna, J. Meijer, I. Schwarz, M. Plenio, A. Retzker, L. P. McGuinness, and F. Jelezko, "Submillihertz magnetic spectroscopy performed with a nanoscale quantum sensor," Science **356**, 832-837 (2017)
[45] J.-H. Kim, S. Aghaeimeibodi, C. J. K. Richardson, R. P. Leavitt, and E. Waks, "Super-Radiant Emission from Quantum Dots in a Nanophotonic Waveguide," Nano Letters **18**, 4734-4740 (2018)